\documentclass[twoside]{article}
\usepackage{amsmath,amssymb,amsfonts}
\pdfoutput=1
\usepackage{graphics}
\DeclareGraphicsExtensions{.pdf, .jpg}
\newtheorem{theorem}{Theorem}[section]

\begin{document}
\title{On the entropy flows to disorder}
\pagestyle{myheadings}
\markboth{On the entropy flows to disorder}{C.T.J. Dodson}


\author{ C.T.J. Dodson \\{\small\it School of Mathematics, University of Manchester,
  Manchester M13 9PL, UK}\\
  {\small\it ctdodson@manchester.ac.uk}
}

\date{\small\it \today}
\maketitle
\begin{abstract}
 Gamma distributions, which contain the exponential as a
special case, have a distinguished place in the representation of
near-Poisson randomness for statistical processes; typically, they represent
distributions of spacings between events or voids among objects.
Here we look at the properties of the
Shannon entropy function and calculate its corresponding flow curves, relating
them to examples of constrained degeneration from ordered processes.
We consider also univariate and bivariate gamma, as well as Weibull distributions
since these include exponential distributions.

\end{abstract}

\section{Introduction}
The smooth family of gamma probability density functions is given by
\begin{equation}\label{gammapdf}
    f: [0,\infty) \rightarrow [0,\infty): x \mapsto
    \frac{e^{-\frac{x \kappa }{\mu }} x^{\kappa -1} \left(\frac{\kappa }{\mu
   }\right)^{\kappa }}{\Gamma (\kappa )} \ \ \ \mu, \kappa > 0.
\end{equation}
Here $\mu$ is the mean, and the standard deviation $\sigma,$
given by $\kappa=(\frac{\mu}{\sigma})^2,$ is  proportional
to the mean. Hence the coefficient of variation $\frac{1}{\sqrt{\kappa}}$
is unity in the case that (\ref{gammapdf}) reduces to
the exponential distribution. Thus, $\kappa=1$ corresponds to an underlying Poisson
random process complementary to the exponential distribution. When $\kappa<1$ the
random variable $X$ represents spacings between events that are
more clustered than for a Poisson process and when  $\kappa>1$ the
spacings $X$ are more uniformly distributed than for Poisson.
The case when $\mu=n$ is a positive integer and $\kappa=2$ gives
the Chi-Squared distribution with $n-1$ degrees of freedom; this
is the distribution of $\frac{(n-1)s^2}{\sigma_G^2}$ for variances $s^2$
of samples of size $n$ taken from a Gaussian population with variance $\sigma_G^2.$

\begin{figure}
\begin{center}
\begin{picture}(300,180)(0,0)
\put(0,0){\resizebox{10cm}{!}{\includegraphics{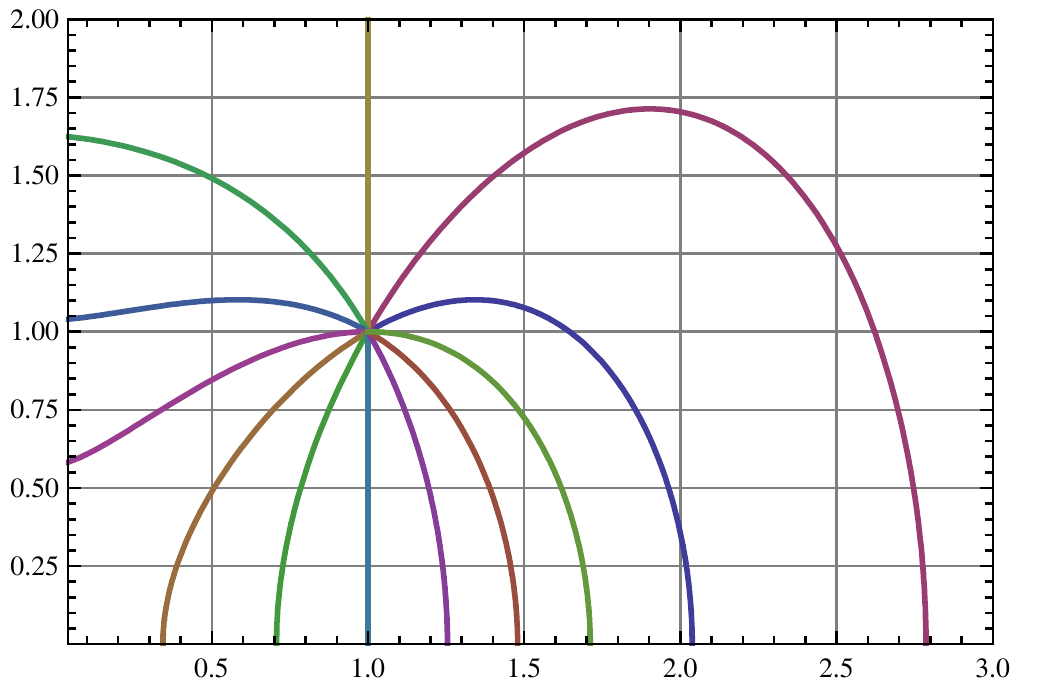}}}
\put(-10,148){$\kappa$}
\put(170,-3){$\mu$}
\end{picture}
\end{center}
\caption{{\em Some examples of maximally extended
geodesics passing through $(\mu,\kappa)=(1,1)$ in the gamma 2-manifold.}}
\label{35gamgeod11}
\end{figure}

 The gamma distribution has a conveniently tractable information
geometry~\cite{AN,InfoGeom}, and the Riemannian metric in the
2-dimensional manifold of gamma distributions (\ref{gammapdf}) is
\begin{eqnarray}
\label{Gammametricgammakappa}
  \left[g_{ij}\right](\mu,\kappa)&=&  = \left[ \begin{array}{cc}
        \frac{\kappa}{{\mu}^2}  &    0  \\
                            0  &   \frac{d^2}{d\kappa^2}\log(\Gamma)-\frac{1}{\kappa}
\end{array} \right].
\end{eqnarray}
So the coordinates $(\mu,\kappa)$ yield an orthogonal basis
of tangent vectors, which is useful in calculations because then
the arc length function is simply
$$ds^2=\frac{\kappa}{\mu^2} \, d\gamma^2 +
        \left(
        \left(\frac{\Gamma'(\kappa)}{\Gamma(\kappa)}\right)' -
        \frac{1}{\kappa}\right)\, d\kappa^2 .$$
The system of geodesic equations is difficult to solve analytically but
numerical solutions using the {\em Mathematica} programs of
Gray~\cite{gray} were obtained in~\cite{InfoGeom}.
Figure~\ref{35gamgeod11} shows a spray of some
 maximally extended geodesics\index{geodesic spray} emanating from
the point $(\mu,\kappa)=  (1,1).$
Geodesic curves have tangent vectors that are parallel along them and
yield minimal information arc length distances in the gamma 2-manifold.
We note the following important uniqueness property:
\begin{theorem}[Hwang and Hu~\cite{hwang}]\label{hwangthm}
For independent positive random variables with a common probability density function $f,$
having independence of the sample mean and the sample coefficient of variation is
equivalent to $f$ being the gamma distribution.
\end{theorem}
This property is one of the main reasons for the large number of applications
of gamma distributions: many near-random natural processes have standard
deviation approximately proportional to the mean~\cite{InfoGeom}. Given a set of
identically distributed, independent data values $X_1,X_2,\ldots ,X_n,$
the `maximum likelihood' or `maximum entropy' parameter values $\hat{\mu}, \hat{\kappa}$
for fitting the gamma distribution  (\ref{gammapdf})
are computed in terms of the mean and mean logarithm of the $X_i$
by maximizing the likelihood function
$$L_f(\mu,\kappa) = \prod_{i=1}^n f(X_i;\mu,\kappa).$$
By taking the logarithm and setting the gradient to zero we obtain
\begin{eqnarray}
\hat{\mu}&=&\bar{X}=\frac{1}{n}\sum^n_{i=1}X_i\\
\log\hat{\kappa} -\frac{\Gamma'(\hat{\kappa})}{\Gamma(\hat{\kappa})}
   & = & \log\bar{X} - \frac{1}{n}\sum^n_{i=1}\log X_i \nonumber \\
   & = & \log\bar{X} - \overline{\log X} . \label{maxlikgam}
\end{eqnarray}

\section{Gamma entropy flows}
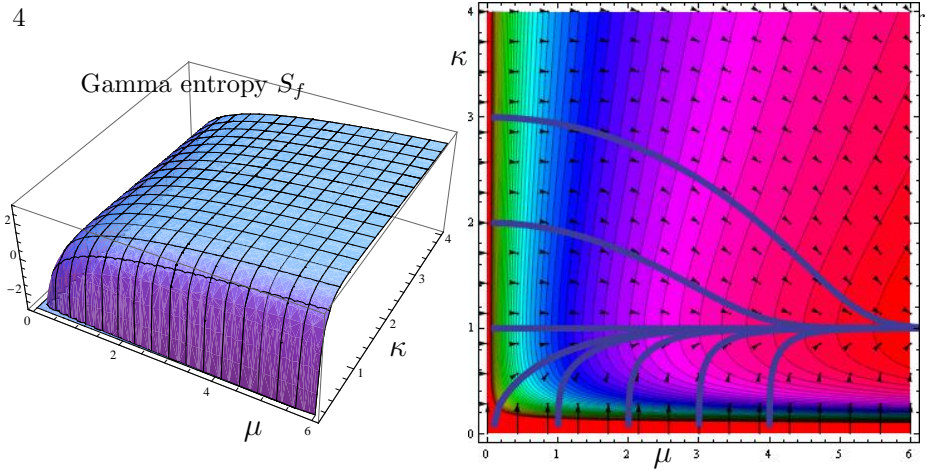
\begin{figure}
\begin{center}
\begin{picture}(300,130)(0,0)
\put(-25,0){\resizebox{6cm}{!}{\includegraphics{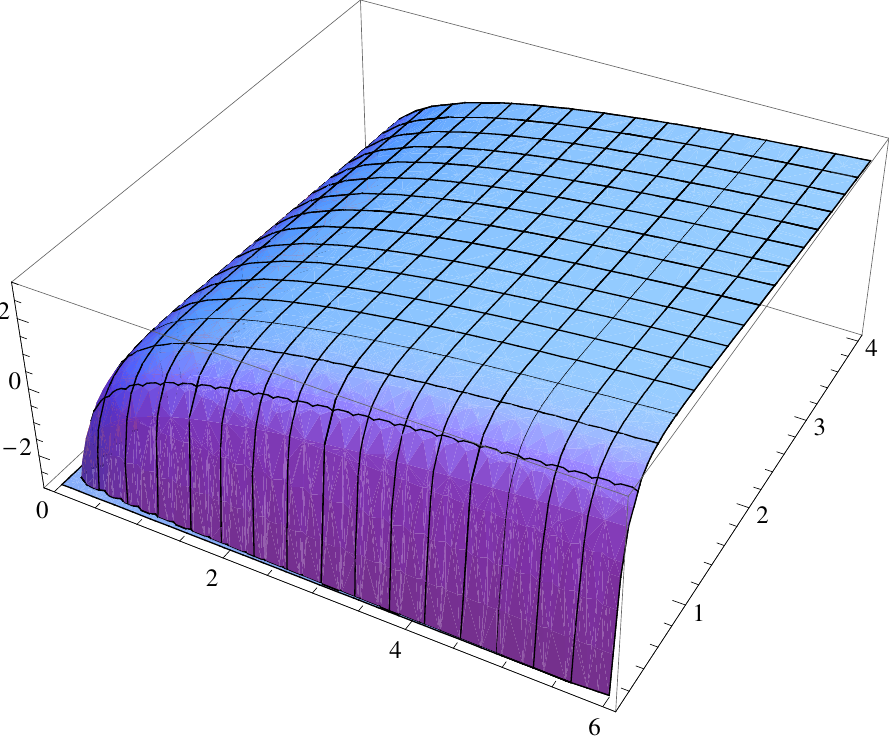}}}
\put(150,-10){\resizebox{6cm}{!}{\includegraphics{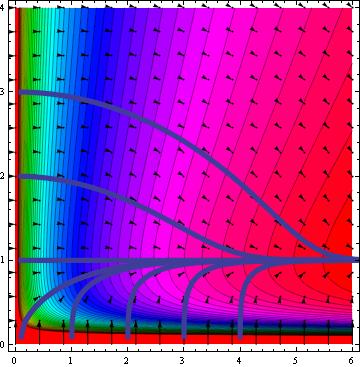}}}
\put(0,130){{ Gamma entropy $S_f$}}
\put(120,30){{\large $\kappa$}}
\put(65,0){\large $\mu$}
\put(143,140){{\large $\kappa$}}
\put(220,-10){\large $\mu$}
\end{picture}
\end{center}
\caption{{\em Shannon entropy function $S_f$ for the
gamma family as a surface with respect to mean $\mu$ and $\kappa$ (left) and
as a contour plot with entropy gradient flow and integral curves (right).
The asymptote is $\kappa=1,$ the exponential case of maximum disorder.}}
\label{GamEntSurfCont}
\end{figure}

Papoulis~\cite{papoulis} Chapter 15 gives an account of the role of the
Shannon entropy function in probability theory, stochastic processes
and coding theory.
The entropy of  (\ref{gammapdf}) is shown in Figure~\ref{GamEntSurfCont} using
\begin{eqnarray}\label{gamentrop}
    S_f &=& -\int_0^\infty f \, \log f \ dx :\mathbb{R}^{2+} \rightarrow \mathbb{R} \nonumber \\
      (\mu,\kappa) &\mapsto&  \kappa -\log \left(\frac{\kappa }
      {\mu }\right)+\log (\Gamma (\kappa ))-(\kappa -1) \psi(\kappa )\\
      {\rm with \ gradient} && \nonumber \\
     \nabla  S_f(\mu,\kappa) &=& \left(\frac{1}{\mu },\
     -\frac{(\kappa -1) \left(\kappa  \psi'(\kappa )-1\right)}{\kappa
   }\right).
\end{eqnarray}
where $\psi= \frac{\Gamma'}{\Gamma}$ is the digamma function. At fixed $\kappa,$
the entropy increases like $\log \mu.$
At fixed mean $\mu$, the maximum entropy is given by $\kappa=1,$
the exponential distribution case of maximal disorder or chaos.

 Figure~\ref{GamEntSurfCont} on the right shows entropy as a
contour plot with superimposed also some examples of
integral curves of the entropy gradient flow field, namely curves $c$ satisfying
\begin{equation}\label{gamintcurv}
    c:[0,\infty)\rightarrow \mathbb{R}^2 : \dot{c}(t)=\nabla {S_f}_{|(c(t))}.
\end{equation}
By inspection, we can see that the entropy gradient components are each in one
variable only and in particular the first component has solution
$$\mu(t)=\mu_0e^t$$
so the mean increases exponentially with time.
Such curves represent typical trajectories for processes subordinate to gamma
distributions; the processes become increasingly disordered as $\kappa \rightarrow 1.$
The entropy gradient curves
correspond to systems with external input---the mean increases
as disorder increases. The asymptote is
$\kappa=1,$ the exponential case of maximum disorder. Conversely,
the reverse direction of the curves corresponds to evolution from total
disorder to other states (clustered for $\kappa<1,$ and smoothed out,
`more crystal-like', for $\kappa>1$) while the mean is allowed to reduce---somewhat
like the situation after the Big Bang, see Dodson~\cite{vpf06}.

\section{Constrained degeneration of order}
Lucarini~\cite{lucarini08} effectively illustrated the degeneration of order
in his perturbations of the simple 3D cubic crystal lattices (SC, BCC, FCC) by an
increasing spatial Gaussian noise. Physically, the perturbing spatial noise
intensity corresponds somewhat to a lattice
temperature in the structural symmetry degeneration.
With rather moderate
levels of noise, quite quickly the three tessellations became indistinguishable.
In the presence of intense noise they all converged to the 3D Poisson-Voronoi tessellations,
for which exact analytic results are known~\cite{finch}. Moreover, in
all cases the gamma distribution was an excellent model for the
observed probability density functions of all metric and topological
properties. See also Ferenc and N\'{e}da~\cite{ferencneda08} for some analytic approximations
using gamma distributions for two and three dimensional Poisson-Voronoi cell
size statistics. Lucarini provided plots showing the evolution of
the mean and standard deviation of these properties
as they converge asymptotically
towards the Poisson-Voronoi case, illustrating the degeneration
of crystallinity from $\kappa\sim\infty$ to lower values.

Of course, the constraint of remaining tessellations, albeit highly disordered ones,
precludes convergence down to the maximum entropy limit $\kappa=1.$ In fact the
limiting values are $\kappa\approx 16$ for number of vertices and the same for
number of edges  and $\kappa\approx 22$ for the
number of faces; actually these are discrete random variables and the gamma is not
appropriate. However, for the positive real random variables, polyhedron volume in the limit has
 $\kappa\approx 5.6$ and polygon face area $\kappa\approx 16.$
Lucarini~\cite{lucarini2d08} had reported similar
findings for the 2D case of perturbations of the three regular tessellations of the plane:
square, hexagonal and triangular. There also the gamma distribution gave a good fit for the
distributions during the degeneration of the regular tessellations to the 2D Poisson-Voronoi
case; the limiting values were   $\kappa\approx 16$ for the perimeter of polygons
 and  $\kappa\approx 3.7$ for areas.

 We can give another perspective on the level of constraint persistent in the limits
 for these disordered tessellations: for infinite random matrices the best fitting
 gamma distributions for eigenvalue spacings have $\kappa=  2.420, \ 4.247, \ 9.606$
 respectively for orthogonal, unitary and symplectic Gaussian ensembles~\cite{InfoGeom}.
 These values in a sense indicate the increasing statistical constraints of algebraic relations
 as the ensembles change through orthogonality, unitarity and symplectivity.
\begin{table}
\begin{center}
\begin{tabular}{|c||c|c|c|c|}
  \hline
  Prime Sequence & $\mu_P$ & $\sigma_P$ & $cv_P=\frac{\sigma_P}{\mu_P}$ & $\kappa_P$
  \\ \hline\hline
         1-100,000 & 13.00 & 10.58 & 0.814 & 1.74\\
   100,000-200,000 & 14.49 & 11.93 & 0.823 & 1.67\\
   200,000-300,000 & 15.05 & 12.48 & 0.830 & 1.67\\
   300,000-400,000 & 15.43 & 12.78 & 0.829 & 1.64\\
   400,000-500,000 & 15.64 & 12.97 & 0.829 & 1.64\\
   500,000-600,000 & 15.88 & 13.23 & 0.833 & 1.62\\
   600,000-700,000 & 16.08 & 13.36 & 0.831 & 1.62\\
   700,000-800,000 & 16.20 & 13.51 & 0.834 & 1.62\\
   800,000-900,000 & 16.35 & 13.59 & 0.831 & 1.61\\
 900,000-1,000,000 & 16.46 & 13.75 & 0.835 & 1.60\\
  1-10,000,000 & 17.81 & 15.01& 0.843 & 1.56\\
  1-100,000,000& 20.07 & 16.97& 0.846 & 1.34\\
  \hline
\end{tabular}
\end{center}
\caption{\em Statistical properties of the spacings between
consecutive prime numbers: mean $\mu_P,$ standard deviation
$\sigma_P,$ coefficient of variation $cv_P=\frac{\sigma_P}{\mu_P},$ maximum likelihood gamma
parameter $\kappa_P,$ for
each of the first ten blocks of 100,000 primes, and the overall data for the first 10 million
primes and the first 100 million primes.}\label{primestats}
\end{table}

 In the context of analytic number theory, gamma distributions give approximate distributions
 for the the reported data on spacings between zeros of the Riemann zeta function~\cite{odlyzko}.
 The best fit gamma distributions to spacings between the first two million
zeros of the Riemann zeta function has $\kappa\approx  5.6$~\cite{InfoGeom}.

\begin{figure}
\begin{center}
\begin{picture}(300,180)(0,0)
\put(0,0){\resizebox{10cm}{!}{\includegraphics{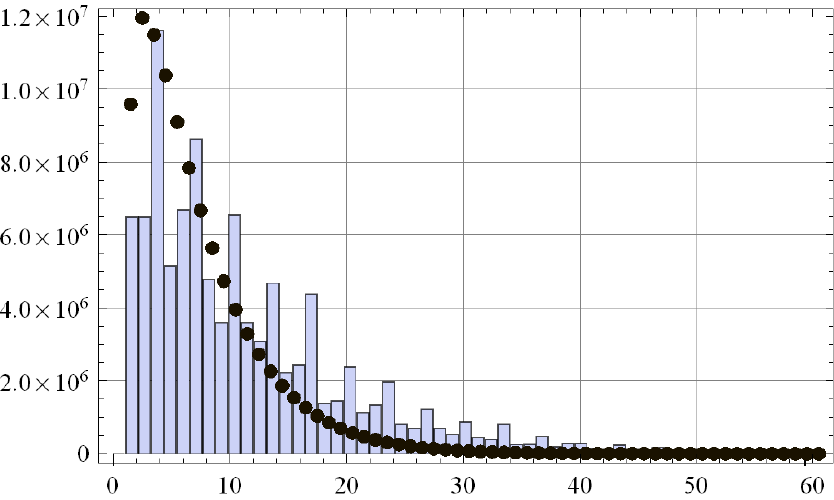}}}
\put(220,-10){Rank order position}
\end{picture}
\end{center}
\caption{{\em Frequency histogram of the spacings between the first 100 million primes
and the maximum likelihood gamma frequencies (points) with $\kappa_P=1.34.$
The early peaks in rank order are at spacings of 6,\, 12,\, 10,\, 18, \,2 and 4.
The mean spacing $\mu_p=20.07$ occurs close to rank order 11. }}
\label{gamfit}
\end{figure}

The spacings between
consecutive prime numbers in successive blocks have surprisingly stable means and standard
deviations, as we see in Table~\ref{primestats}. This gives the mean $\mu_P,$ standard deviation
$\sigma_P,$ coefficient of variation $cv_P,$ maximum likelihood gamma parameter $\kappa_P$ for
each of the first ten blocks of 100,000 primes, and the same data for the first 100 million primes.
Of course, the gamma distribution is not a good fit, as we can see in Figure~\ref{gamfit}
for the first 100 million primes. From Table~\ref{primestats}, the mean drifts up with
the local mean size, as we would expect from the Prime Number Theorem, but
so also does the standard deviation in proportion, hence keeping the coefficient of variation
nearly constant. This stability suggests that there should be some qualitative number
theoretic property which is being reflected and that the distribution of spacings among
these early primes is somewhat close to a Poisson process ($\kappa=1$).

 For the SARS disease epidemic outbreak~\cite{BL,who} the gamma distribution gave a good
model and the infection process was approximated by $\kappa\approx  3,$ cf.~\cite{stochrateproc}.

\section{Bivariate gamma processes}
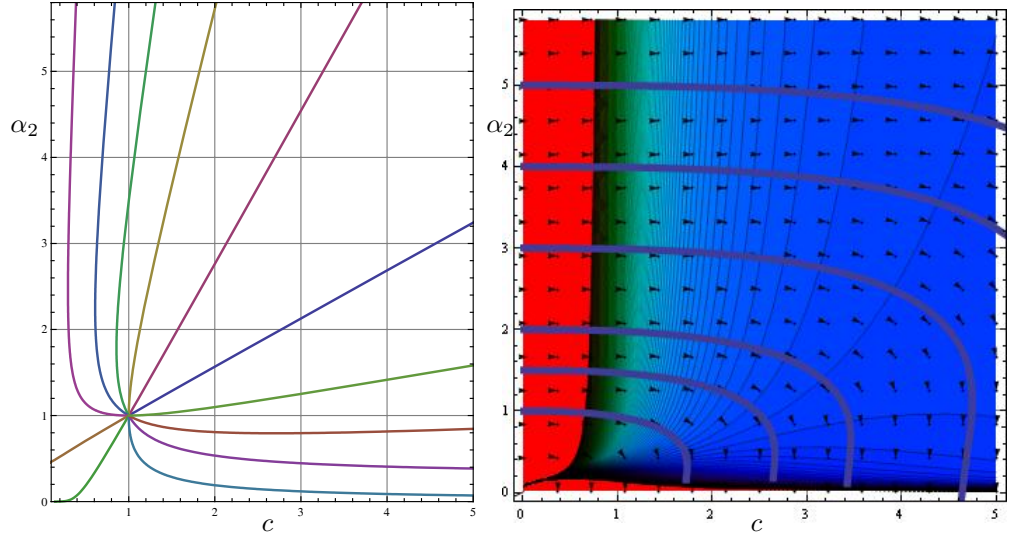
\begin{figure}
\begin{center}
\begin{picture}(300,190)(0,0)
\put(-20,0){\resizebox{6cm}{!}{\includegraphics{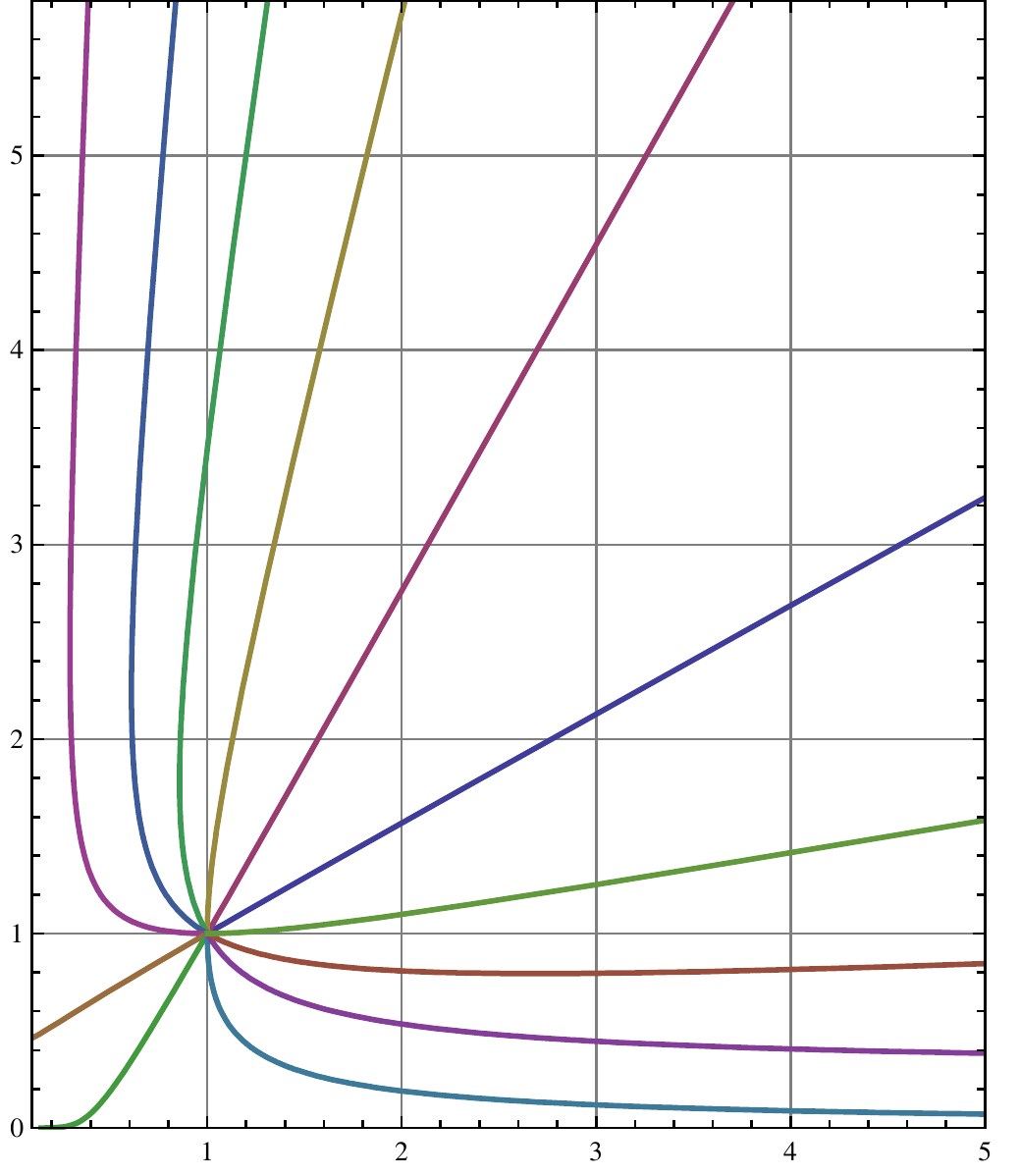}}}
\put(156,0){\resizebox{6.7cm}{!}{\includegraphics{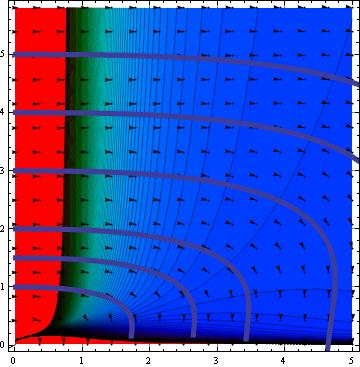}}}
\put(-30,148){$\alpha_2$}
\put(65,-3){$c$}
\put(150,148){$\alpha_2$}
\put(250,-3){$c$}
\end{picture}
\end{center}
\caption{{\em Geodesics passing through $(c,\alpha_2)=(1,1)$
 (left) and a contour plot of the entropy
with some integral gradient curves (right), in the McKay submanifold
$M_1$ which has $\alpha_1=1$.}}\label{4McKayM1geod}
\end{figure}

Next we consider the bivariate case of a pair of coupled gamma processes.
The McKay family can be thought of as giving the probability density for the two
random variables $X$ and $Y=X+Z$ where $X$ and $Z$ both have gamma distributions.
This smooth bivariate family $M$ of density functions
is defined in the positive octant of random variables
$ 0<x<y<\infty $ with parameters $\alpha_{1},c,\alpha_{2}\in\mathbb{R}^+$
and probability density functions
\begin{eqnarray}
m(x,y) =
\frac{c^{(\alpha_{1}+\alpha_{2})}x^{\alpha_{1}-1}(y-x)^{\alpha_{2}-1}
e^{-c y}}{\Gamma(\alpha_{1})\Gamma(\alpha_{2})}\ . \label{mckayaca}
\end{eqnarray}
The marginal density functions, of $X$ and $Y$
are:
\begin{eqnarray}
 m_{X}(x) &= & \frac{c^{\alpha_{1}} x^
{\alpha_{1}-1}e^{-c\,x}}{\Gamma(\alpha_{1})},\quad x
>0 \\
 m_{Y}(y)&=&\frac{c^{(\alpha_{1}+\alpha_{2})}
y^{(\alpha_{1}+\alpha_{2})-1}e^ {-c\,
y}}{\Gamma(\alpha_{1}+\alpha_{2})}, \quad y>0.
\end{eqnarray}
Note that we cannot have both marginal distributions exponential.
The covariance and correlation coefficient of $X$ and $Y$ are:
$$ \sigma_{12}=\frac{\alpha_{1}}{c^{2}}   \ \ {\rm and}  \ \
 \rho(X,Y)= \sqrt{\frac{\alpha_{1}}{\alpha_{1}+\alpha_{2}}}. $$
Unlike other bivariate gamma families, the McKay information geometry is surprisingly tractable
and there are a number of applications discussed in Arwini and Dodson~\cite{InfoGeom}.
In fact the parameters  $\alpha_{1},c,\alpha_{2}\in\mathbb{R}^+$ are natural coordinates for
the 3-dimensional manifold $M$ of this family. The Riemannian information metric is
\begin{eqnarray}
   [g_{ij}(\alpha_{1},c,\alpha_{2})]=\left[ \begin{array}{ccc}
 \psi'({\alpha }_1)& -\frac{1}{c} & 0 \\
  - \frac{1}{c} & \frac{{{\alpha }_1} + {{\alpha }_2}}{c^2}& - \frac{1}{c} \\
  0 & -\frac{1}{c} & \psi'({\alpha }_2)
\end{array} \right] .\label{3mckaymetric}
\end{eqnarray}
It is difficult to present graphics of curves in the McKay 3-manifold $M,$ but it has
an interesting 2-dimensional submanifold $M_{1}\subset M$: $\alpha_{1}=1.$ The density
functions are of form:
 \begin{eqnarray}
h(x,y;1,c,\alpha_{2}) = \frac{c^{1+\alpha_{2}}(y-x)^{\alpha_{2}-1}
e^{-c\,y}}{\Gamma(\alpha_{2})} \ ,\label{m1}
\end{eqnarray}
defined on $ 0<x<y<\infty $ with parameters $ c,\alpha_{2}\in\mathbb{R}^+.$ The
correlation coefficient and marginal functions of $X$ and $Y$ are
given by:
\begin{eqnarray}
 \rho(X,Y)&=&\frac{1}{\sqrt{1+\alpha_{2}}}  \\
 h_{X}(x) &= & c\, e^{-c\,
x},\quad x
>0 \\
 h_{Y}(y)&=&\frac{c^{(1+\alpha_{2})}
y^{\alpha_{2}}e^ {-c\, y}}{\alpha_{2}\,\Gamma(\alpha_{2})}, \quad
y>0
\end{eqnarray}
In fact $ \alpha_{2}=\frac{1-\rho^2}{\rho^2},$ which in
applications would give a measure of the variability not due to the
correlation.
The matrix of metric components $[g_{ij}]$ on $M_1$ is
\begin{eqnarray}
 [g_{ij}]=\left[ \begin{array}{ccc}
 \frac{1 + {{\alpha }_2}}{c^2}& -\frac{1}{c} \\
  -\frac{1}{c}& \psi'({\alpha }_2)
\end{array} \right] \ .
\label{1mckaymetric}
\end{eqnarray}
Some geodesics emanating from $(\alpha_2,c)=(1,1)\in M_1$ are shown
on the left of Figure~\ref{4McKayM1geod}.
The density functions can be presented also in terms of the positive parameters
$(\alpha_{1},\sigma_{12},\alpha_{2})$ where $\sigma_{12}$ is the covariance of $X$ and $Y$
\begin{eqnarray}
m(x,y;\alpha_{1},\sigma_{12},\alpha_{2}) &=&
\frac{(\frac{\alpha_{1}}{\sigma_{12}})^{\frac{(\alpha_{1}+\alpha_{2})}{2}}x^
{\alpha_{1}-1}(y-x)^{\alpha_{2}-1}
e^{-\sqrt{\frac{\alpha_{1}}{\sigma_{12}}}y}}{\Gamma(\alpha_{1})\Gamma(\alpha_
{2})} \label{masiga}\\
 m_{X}(x) &= & \frac{(\frac{\alpha_{1}}{\sigma_{12}})^{\frac{\alpha_{1}}{2}} x^
{\alpha_{1}-1}e^{-\sqrt{\frac{\alpha_{1}}{\sigma_{12}}}
x}}{\Gamma(\alpha_{1})},\quad x
>0 \\
 m_{Y}(y)&=&\frac{(\frac{\alpha_{1}}{\sigma_{12}})^{\frac{(\alpha_{1}+\alpha_{2})}{2}}
y^{(\alpha_{1}+\alpha_{2})-1}e^
{-\sqrt{\frac{\alpha_{1}}{\sigma_{12}}}
y}}{\Gamma(\alpha_{1}+\alpha_{2})}, \quad y>0 .
\end{eqnarray}
\begin{figure}
\begin{center}
\begin{picture}(300,140)(0,0)
\put(-25,-20){\resizebox{7cm}{!}{\includegraphics{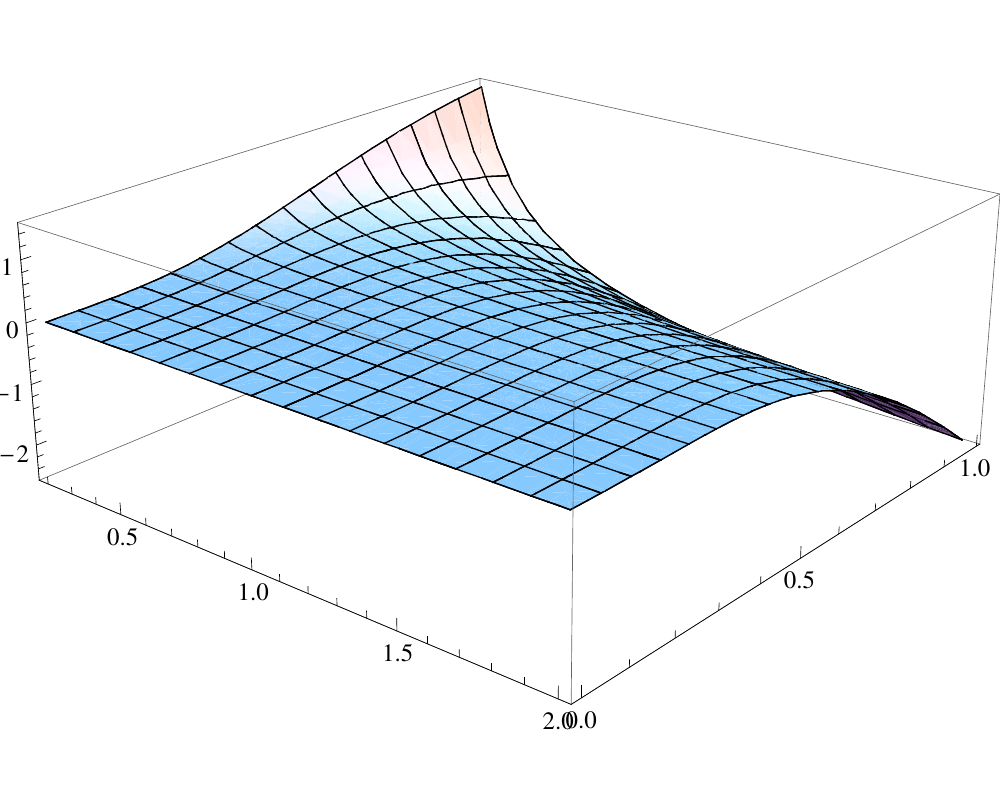}}}
\put(180,-10){\resizebox{5cm}{!}{\includegraphics{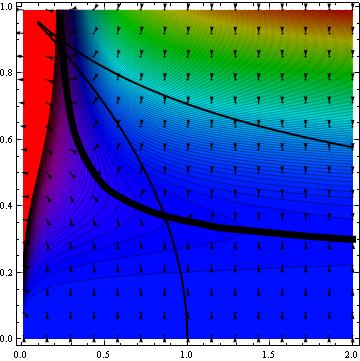}}}
\put(30,130){{\large McKay entropy ${S_m}_{|M_1}$}}
\put(120,10){{$\rho$}}
\put(30,10){ $\alpha_2$}
\put(170,120){{$\rho$}}
\put(300,-15){ $\alpha_2$}
\end{picture}
\end{center}
\caption{{\em Surface representation of the Shannon entropy function $S_m$ for the
submanifold $M_1$ of the McKay family with respect to parameter $\alpha_2$ and
correlation coefficient $\rho$ (left), and contour plot (right).
Superimposed also are gradient flow arrows and the thick curve is
the locus of maximum entropy.
The two thinner curves show the loci of $\alpha_1=1$
(upper curve) and
$\alpha_1+\alpha_2=1$ (lower curve),
 which correspond, respectively, to Poisson random processes for the
$X$ and $Y$ variables. }}
\label{MEntSurf}
\end{figure}
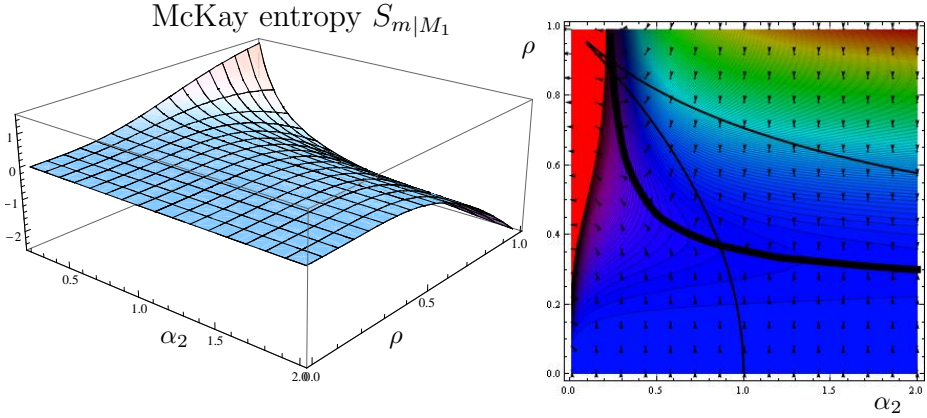
The entropy function is
\begin{eqnarray}
  S_m &:&\mathbb{R}^{3+} \rightarrow \mathbb{R} \nonumber \\
  (\alpha_{1},c,\alpha_{2})&\mapsto &\sqrt{\alpha _1}c^{-\alpha _1-1}  K\\
  K&=&
 \log\frac{c^2}{\Gamma(\alpha _1)\Gamma(\alpha _2)}+(\alpha _1-1) \psi(\alpha _1)
 +(\alpha _2-1) \psi(\alpha _2)-(\alpha _1+\alpha_2).\nonumber\\
  {\rm On} \ M_1  && \nonumber \\
   {S_m}_{|M_1} &=& \frac{1}{c^2}\left(\log\frac{c^2}{\Gamma(\alpha _2)}+
   (\alpha _2-1) \psi(\alpha _2) -(1+\alpha_2)\right)\\
   \nabla {S_m}_{|M_1}&=&\left(
   \frac{2}{c^3} \left(\log \left(\frac{\Gamma (\alpha _2)}{c^2}\right)
   -\psi(\alpha _2) (\alpha _2-1)+\alpha _2+2\right),
    \frac{1}{c^2}\left(\psi'(\alpha _2) (\alpha _2-1)-1\right)
                              \right).\nonumber
\end{eqnarray}
On the right of Figure~\ref{4McKayM1geod} is a contour
plot of the entropy showing its gradient field and some integral curves.
It may be helpful to express the entropy in terms of $\alpha_2$ and $\rho$, which gives
\begin{eqnarray}
  {S_m}_{|M_1} &=& \frac{4 \rho ^4 }{\left(\rho ^2+1\right)^2}
  \left(\log \frac{\left(\rho ^2+1\right)^2}
  {4 \rho ^4  \left(\Gamma \left(\alpha _2\right)\right)}
    + \left(\alpha _2-1\right)\psi\left(\alpha _2\right)
  -(1+\alpha _2)\right) .
\end{eqnarray}
The McKay entropy in $M_1$ with respect to $\alpha_2,\rho$
is shown in  Figure~\ref{MEntSurf} (left) with the gradient flow on
a contour plot (right) together with the approximate locus curve of maximum entropy.
The two thinner curves show the loci of $\alpha_1=1$
(upper curve) and
$\alpha_1+\alpha_2=1$ (lower curve),
 which correspond, respectively, to Poisson random processes for the
$X$ and $Y$ variables.

 Qualitatively, what we may see in  Figure~\ref{MEntSurf} is that for correlated
random variables $X$ and $Y$ subordinate to the bivariate gamma density
(\ref{m1}), the maximum entropy locus is roughly hyperbolic.
The maximum entropy curve is rather insensitive to
the correlation coefficient $\rho$ when $\alpha_2>1$ and the difference $Y-X$
is dispersed more evenly than Poisson. When $Y-X$ is actually Poisson random, with $\alpha_2=1,$
the critical value is at $\rho\approx 0.355.$
However, as $\alpha_2$ reduces further---corresponding to clustering of
$Y-X$ values---so the locus turns rapidly to increasing correlation.
If we take the situation of a bivariate gamma process with constant marginal
mean values, then the McKay probability density has constant correlation coefficient $\rho;$
in this case the gradient flow lies along lines of constant $\alpha_2.$

 Dodson~\cite{stochrateproc} gives an application to an epidemic model in which the latency
and infectivity for individuals in a population are jointly distributed
properties controlled by a bivariate gamma distribution.
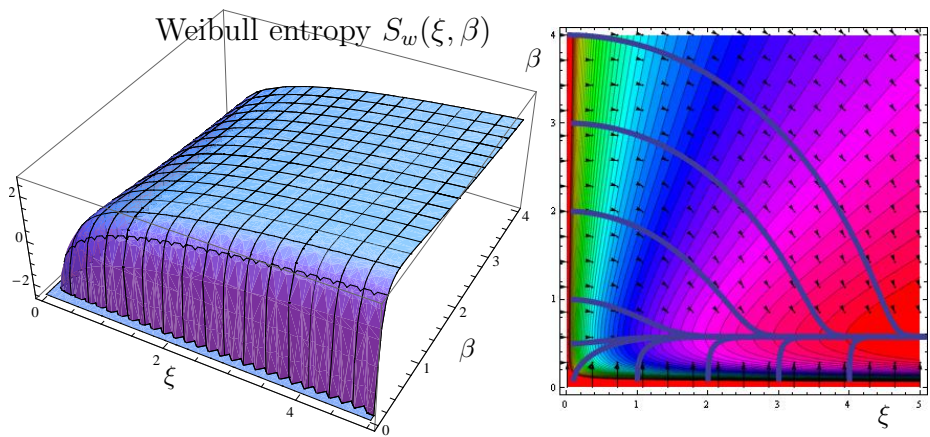
\begin{figure}
\begin{center}
\begin{picture}(300,140)(0,0)
\put(-25,-20){\resizebox{7cm}{!}{\includegraphics{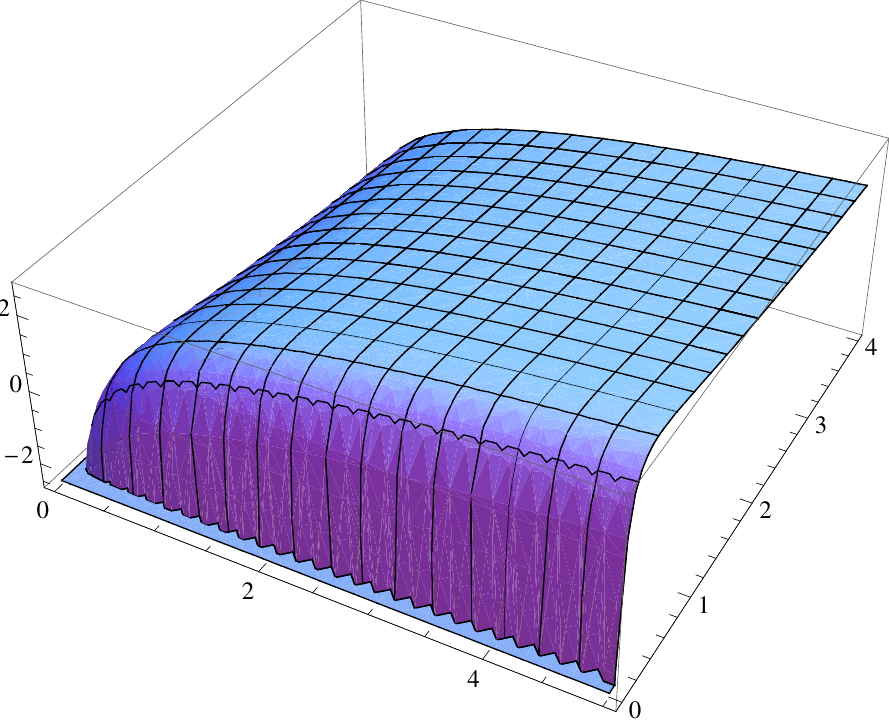}}}
\put(180,-10){\resizebox{5cm}{!}{\includegraphics{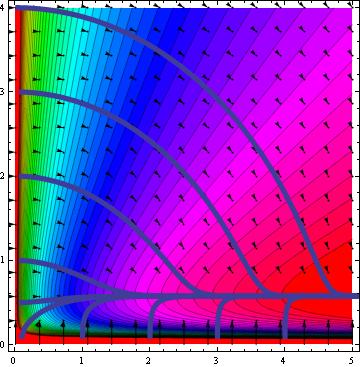}}}
\put(30,130){{\large Weibull entropy ${S_w}(\xi,\beta)$}}
\put(145,10){{$\beta$}}
\put(30,0){ $\xi$}
\put(170,120){{$\beta$}}
\put(300,-15){ $\xi$}
\end{picture}
\end{center}
\caption{{\em Surface representation of the Shannon entropy function $S_w$ for the
Weibull family (left), and contour plot (right) with gradient flow and integral curves. }}
\label{WEntSurfFlow}
\end{figure}
\section{Weibull processes}
Like the gamma family, the Weibull family of distributions
contains the exponential distribution as a special case;
it has wide application in models for reliability and
lifetime statistics for random variable $t>0.$ The probability density
function can be presented in terms of positive parameters $\xi,\beta$
\begin{eqnarray}\label{w}
    w: \mathbb{R^+} &\rightarrow& \mathbb{R} :
    t \mapsto \frac{\beta}{\xi }\, \left(\frac{t}{\xi }\right)^{\beta -1}
    e^{-\left(t/\xi\right)^{\beta }}.
\end{eqnarray}
In applications of (\ref{w}), reliability $R(t)$ is the probability of
survival to time $t$ and it is related to the failure
rate $Z(t)$ at time $t$ through
\begin{equation}
  R(t) = \int_t^\infty \, w(t)\, dt ={ e^{-(t/\xi)}}^{\beta } \ {\rm and} \
  Z(t) = \frac{w(t)}{R(t)} = \beta \left(\frac{1}{\xi}\right)^\beta \ t^{\beta -1}.
\end{equation}
The Weibull mean, standard deviation and entropy are
\begin{eqnarray}\label{wmsdE}
    \mu_w &=& \xi \, \Gamma(1+\frac{1}{\beta}) \label{wm}\\
   \sigma_w &=& \xi \, \sqrt{\Gamma \left(\frac{\beta +2}
   {\beta }\right)-\Gamma \left(1+\frac{1}{\beta }\right)^2}\label{wsd}\\
   S_w(\xi,\beta) &=& -\log (\beta )-\log \left(\frac{1}{\xi }\right)-\frac{\gamma }{\beta
   }+\gamma +1 \label{wE}\\
   \nabla S_w(\xi,\beta) &=& (\frac{1}{\xi}, \, \frac{\gamma-\beta}{\beta^2}) \label{gradSw}.
\end{eqnarray}
In (\ref{wE}), $\gamma$ is the Euler constant, of value approximately $0.577.$

In case $\beta=\frac{1}{n}$ for positive integer $n,$ then
the coefficient of variation is
$$\frac{\sigma_w}{\mu_w}=\sqrt{\frac{(2n)!}{(n!)^2}-1},$$
 and we see that
the case $\beta=1$ reduces (\ref{w}) to the exponential density with $\mu_w=\sigma_w=\xi$
and hence unit coefficient of variation.
Figure~\ref{WEntSurfFlow} shows a surface plot of the Weibull entropy ${S_w}$ and a corresponding
contour plot with gradient flow and some integral curves. There is a resemblance to the gamma distribution entropy in Figure~\ref{GamEntSurfCont} and from (\ref{gradSw}) again here we have $\mu_w(t)=\mu_w(0)e^t,$ but in the Weibull case the
 asymptotic curve has $\beta=\gamma\approx 0.577,$ which corresponds
to a process with coefficient of variation $\approx 4.65$ compared with unity for the exponential distribution and Poisson randomness.

\subsection*{Acknowledgements}
The author is grateful to V. Lucarini for discussions concerning the
interpretation of his simulations and to W.W. Sampson for help with optimizing
code for analyzing primes.


\end{document}